\documentstyle[multicol,pre,aps,epsf]{revtex}

\begin{document}   

\draft				     

\title{Synchronization Defects and Broken  Symmetry in Spiral Waves}
\author{Andrei Goryachev$^1$, Hugues Chat\'e$^2$,  and Raymond Kapral$^1$}
\address{$^1$Chemical Physics Theory Group, Department of
Chemistry, University of Toronto, Toronto, ON M5S 3H6, Canada\\
$^2$CEA --- Service de Physique de l'Etat Condens\'e, 
Centre d'Etudes de Saclay, 91191 Gif-sur-Yvette, France}
\maketitle
\begin{abstract}
Spiral waves are investigated in oscillatory media
exhibiting period-doubling bifurcations. In the period-doubled and
chaotic regimes, the rotational symmetry of the spiral
wave is broken. The loss of symmetry takes the form of
synchronization defect lines where the phase of the local oscillation 
changes by multiples of $2\pi$. The internal structure and 
geometry of these synchronization defects is studied and a discussion 
of the possible types of defect lines is presented. 
\end{abstract}

\pacs{82.20.Wt, 05.40.+j, 05.60.+w, 51.10.+y}

\begin{multicols}{2} 

\narrowtext

Spatially-distributed oscillatory media may undergo bifurcations where
the period of the orbit doubles at almost every point in the system. 
In the simple oscillatory regime (period-1), before such 
period-doubling bifurcations occur, the system may support stable 
rotationally-symmetric spiral wave solutions. \cite{spexp} The dynamics in 
this regime is typically described by the complex Ginzburg-Landau (CGL) 
equation, the generic equation for an oscillatory medium near the Hopf 
bifurcation point at which oscillations appear \cite{kur}. 
Spiral waves are a well-known feature of this regime where they have been
intensively studied \cite{spiral}. However, they can also exist in media 
where the local dynamics is complex-periodic or chaotic \cite{kle,chate,PRE}.
In this Letter, we investigate the consequences of period-doubling 
bifurcations on the structure and dynamics of spiral waves. We find that
the symmetry of the spiral wave is broken by defect lines 
where the phase of the oscillation changes by multiples of $2\pi$, and we study
the nature of these synchronization defect lines.

We consider the dynamics of spatially-distributed systems governed by  
reaction-diffusion equations of the form
\begin{equation}
\label{rds}
{\partial {\bf c}({\bf r},t) \over \partial t} = {\bf R}({\bf c}({\bf r},t)) 
+ D \nabla^2 {\bf c}({\bf r},t)\;,
\end{equation}		   
where ${\bf c}({\bf r},t)$ is a vector of local concentrations,
$D$ is the diffusion coefficient and
${\bf R}({\bf c}({\bf r},t)$ describes the local reaction kinetics. 
While the phenomena we investigate should be observable for 
any spatially-distributed system exhibiting period-doubling, we 
have considered cases where the spatially-homogeneous
system $\dot{ {\bf c}}(t)  = {\bf R}({\bf c}(t))$ itself exhibits
period-doubling bifurcations. Specifically, the calculations
described here were carried out on the R\"{o}ssler model, $R_x=-c_y-c_z, \;
R_y=c_x + Ac_y, \; R_z=c_x c_z - C c_z + B$, which is well known to exhibit
chaos arising from a cascade of subharmonic bifurcations. 

Spiral waves were initiated as in \cite{chate,PRE}, taking advantage of the 
cyclic character of the projection of the R\"ossler attractor on the $(c_x,c_y)$ 
plane. Various values of the parameter $C$ in the interval [2.5,6.0]
were considered, while
the other parameters were fixed at $A=0.2$ and $B=0.2$. 
The scaled diffusion coefficient was
$D\Delta t/(\Delta r)^2=1.6\times 10^{-2}$  ($\Delta t = 10^{-2}$)
in all calculations.
Simulations were carried out on a disk-shaped domain of radius 255 with no-flux
boundary conditions.
In the period-1 regime, the medium supports a single, stable, 
one-armed spiral wave. As the parameter $C$  increases,
the local dynamics undergoes a period-doubling bifurcation at 
$C=C^*\approx 3.03$. (The spatially-homogenous R\"ossler model bifurcates at 
$C^*_{\rm ode} \approx 2.83$). 
In order to quantitatively characterize this transition, it is convenient
to introduce a local order parameter $\delta c_z({\bf r})$, 
defined as the difference between two
successive maxima of the time series $c_z({\bf r},t)$.
\begin{figure}[htbp]
\begin{center}
\leavevmode
\epsffile{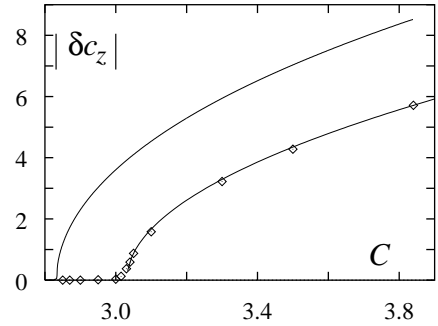}
\end{center}
\caption{Bifurcation diagram, constructed at $r_0 = 130$,
showing the first period-doubling bifurcation 
in the medium supporting a single spiral wave (diamonds). For
comparison, the upper left curve is the bifurcation diagram for a spatially 
homogeneous system.}
\label{fig1}
\end{figure}
Figure~\ref{fig1} shows the dependence of $\delta c_z(r_0)$
on $C$ calculated at a typical point at radius $r_0=130$, 
sufficiently far from both the boundary and the core of the spiral. 
One observes a parabolic increase in $\delta c_z(r_0)$ 
beyond $C^*\approx 3.03$,
typical of a supercritical period-doubling bifurcation. This behavior
is seen at almost every point in the medium (the structure is difficult 
to resolve deep in the core region). 
Along almost any ray emanating from the core, the magnitude of 
$\delta c_z(r)$ varies as $A(r)\sqrt{C-C^*}$ where $A(r)$ behaves 
like the amplitude profile of the spiral itself,
i.e., $A(r) \simeq r^\alpha$ in the core region and is constant elsewhere 
except near the boundary.

\begin{figure}[htbp]
\begin{center}
\leavevmode
\epsffile{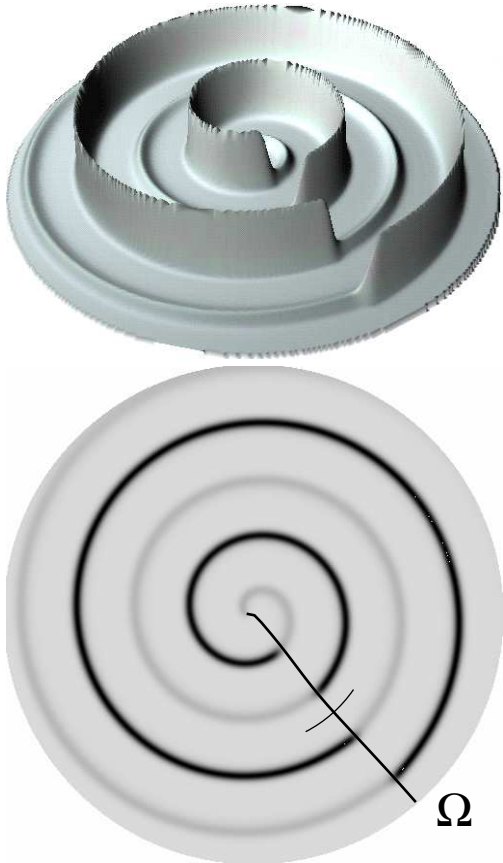}
\end{center}
\caption{Spiral wave in a medium with period-2 local dynamics
at $C=3.84$. Concentration field $c_z({\bf r},t_0)$ is shown as 
elevation in the upper panel and as grey shades in the lower panel. 
The solid line depicts the $\Omega$ curve. The arc
segment at radius $r_0=130$ along which points were taken to 
construct Figs.~\protect\ref{fig3}-\protect\ref{fig5} is also shown.}
\label{fig2}
\end{figure}

\begin{figure}[htbp]
\begin{center}
\leavevmode
\epsffile{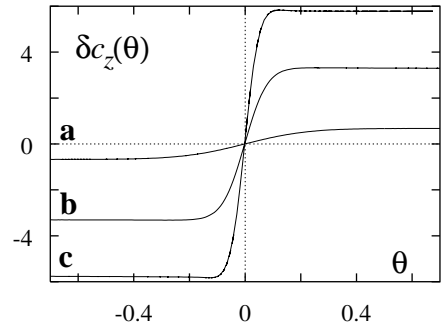}
\end{center}
\caption{Order parameter $\delta c_z(\theta)$ profile across the
$\Omega$ curve  along the arc at radius $r_0=130$ 
indicated in the lower panel of Fig.~\protect \ref{fig2} for three
values of $C$: (a) $C=3.04$, (b) $C=3.10$, (c) $C=3.84$.
Curve (c) is calculated for the medium in Fig.~\protect \ref{fig2}.}
\label{fig3}
\end{figure}

Within the period-2 domain, the spiral wave
acquires a global structure different from that
in a simple periodic medium. Figure~\ref{fig2} shows
the $c_z({\bf r},t_0)$ concentration field at a single time instant
$t_0$. The alternation of high and low $c_z({\bf r},t_0)$ 
maxima unambiguously demonstrates 
that period-2 local temporal dynamics manifests itself 
in the formation of a period-doubled
spiral waveform with broken rotational symmetry
whose wavelength is twice that of the original spiral wave 
in the period-1 medium. 
The lower panel of Fig.~\ref{fig2} shows $c_z({\bf r},t_0)$ 
in grey shades and indicates the curve
connecting the spiral core and the boundary, denoted as $\Omega$, 
where sharp changes in the concentration occur. 
The $\Omega$ curve plays a central role in the organization of the spiral
wave, and its character will be examined below.
The fact that the local dynamics is almost everywhere period-2 
endows the spiral with some unusual features: 
unlike a period-1 medium where the 
concentration is periodic with the period of the spiral rotation, 
here, after one turn of the spiral,
the high and low maxima interchange and it is only after 
two spiral periods that the 
concentration field is restored to its initial value. 
Although the spiral rotates, the $\Omega$ curve is, up to
numerical accuracy, stationary.
In the asymptotic regime, after transients depending on the initial condition,
the shape of the 
$\Omega$ curve takes the form of a straight line segment with a short 
curved portion lying inside the core region.
Figure~\ref{fig3} shows $\delta c_z(\theta)$ for three values of $C$
as function of the polar angle $\theta$ along the arc segment that
crosses the $\Omega$ curve. 
To a very good approximation, the numerical data show that 
$\delta c_z(r,\theta,C)$ varies like
\begin{equation}
\delta c_z(r,\theta,C) \simeq A(r)\sqrt{C \! - \! C^*}\tanh [\kappa r 
(\theta \! - \! \theta_\Omega)(C \! - \! C^*)]
\end{equation}
where $\theta_\Omega$ is the angular position of the $\Omega$ curve 
and $\kappa$ is a 
numerical factor. The  $\Omega$ curve is thus a localized
object whose width remains
approximately constant with $r$ and varies like $1/(C-C^*)$.
As $C$ approaches $C^*$ from above, the width increases until it becomes
comparable to $2\pi$ and the $\Omega$ curve ceases to be a well-defined 
object \cite{localized}.
We now show that the $\Omega$ curve corresponds
to a one-dimensional synchronization defect across which 
the phase of oscillation changes by some multiple of $2\pi$.
One can always uniquely parametrize an orbit of a period-1 oscillation 
by a phase variable $\varphi \in [0,2\pi)$  and 
describe the spatially-distributed medium 
by a phase field $\varphi({\bf r},t)$.
At the center of the spiral lies  a point defect of 
this field characterized by a topological charge 
${1\over 2\pi} \oint \nabla \varphi({\bf  r},t)  \cdot  d{\bf l}=n_{\rm t}$
\cite{merm},
where the integral is taken along a closed curve encircling the defect. 
There exist several possible ways to extend the definition of phase 
for complex-periodic oscillations \cite{pick}. 
We assume that the phase $\phi (t)$ of an $n$-periodic
oscillation is a scalar function of time which increases monotonically by 
$2\pi n$ for each full period of the oscillation $T_n$. 
For some systems, 
$\phi$ can be defined in terms of an angle variable in a suitably-chosen 
coordinate frame in phase space. For the R\"{o}ssler model 
we use a cylindrical coordinate system $(\rho,\varphi,z)$ 
with origin at the unstable fixed point of the spatially-homogenous
system and $z$ along $c_z$.
The phase variable $\phi \in [0,2\pi n)$ then takes the form
$\phi = \varphi + 2\pi m$ where $m$ is an integer with
values from 0 to $n-1$.
While $\varphi$ is a single-valued function of the original dynamical 
variables, the phase $\phi$ of a 
complex-periodic oscillation is not an observable, since 
a knowledge of the entire local orbit is required in order to calculate $m$.
\begin{figure}[htbp]							       
\begin{center}
\leavevmode
\epsffile{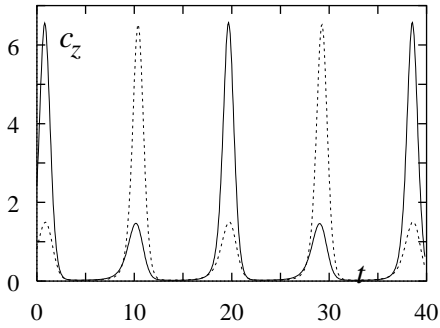}
\end{center}
\caption{Two $c_z(t)$ concentration time series calculated for the medium 
shown in Fig.~\protect\ref{fig2} at points 
${\bf r}_{1,3} = (r_0=130, \theta_{\Omega}\pm 0.05)$ 
on opposite sides of the $\Omega$ curve.}
\label{fig4}
\end{figure}
The $\Omega$ curve can be viewed as a one-dimensional defect of the 
$\phi({\bf r},t)$ field. Figure~\ref{fig4} shows two $c_z(t)$
concentration time series at nearby spatial points on either side 
of the $\Omega$ curve. The two oscillations are shifted relative 
to each other by half a period.
For a period-2 oscillation, this corresponds to a phase shift of 
$\delta \phi = 2\pi$.
The necessity for such a phase synchronization slip in a one-armed
spiral in a period-doubled medium can be understood from the following 
argument. Consider a contour integral of the phase gradient 
$\nabla \phi({\bf r},t)$ taken along a closed loop
$\Gamma$ which surrounds the core of the spiral wave. 
For an arbitrary point ${\bf r}_0 \in \Gamma$, this loop
is just a path in the medium which starts at ${\bf r}_0$,
described by the local $n$-periodic oscillation phase 
$\phi({\bf r},t)$, and returns to the 
same point with same oscillation phase $\phi({\bf r},t)$. 
Thus, the  integral  may take only values $2\pi nk$ equal to multiples
of the full-period phase increment. Since the
phase field is given by 
$\phi ({\bf r},t)=\varphi ({\bf r},t) +2\pi m({\bf r},t)$, we find
\begin{equation}
\label{balance}
\oint \! \nabla \phi ({\bf r},t) \! \cdot \! d{\bf l} = \oint \! \nabla 
\varphi ({\bf r},t) \! \cdot \! d{\bf l}
+2\pi \! \oint \! \nabla m({\bf r},t) \! \cdot \! d{\bf l}\; .
\end{equation}
Simulations for a one-armed spiral show that, at any given time $t$, 
$\oint \nabla \varphi ({\bf r},t) \cdot d{\bf l} = \pm 2\pi$,
regardless of the periodicity of the local dynamics \cite{one-arm}.
Given that the full-period phase increment is $2\pi$ only in the case 
of period-1 dynamics with $m({\bf r},t)\equiv 0$, 
for period-doubled dynamics the integration of $\nabla m({\bf r},t)$
along $\Gamma$ must yield a non-zero contribution to balance (\ref{balance}). 
Since $m({\bf r},t)$ is an  integer function, 
its value changes discontinuously with time and space so that
$\nabla m({\bf r},t)$ is different from zero only at a single 
point: the intersection of $\Gamma$ with the $\Omega$ curve. 
For a  period-2 medium, the addition of a $2\pi$ phase jump  on the 
$\Omega$ curve and a $2\pi$ contribution from integration of  
$\varphi({\bf r},t)$ yields  the necessary full-period increment of $4\pi$.

\begin{figure}[htbp]
\begin{center}
\leavevmode
\epsffile{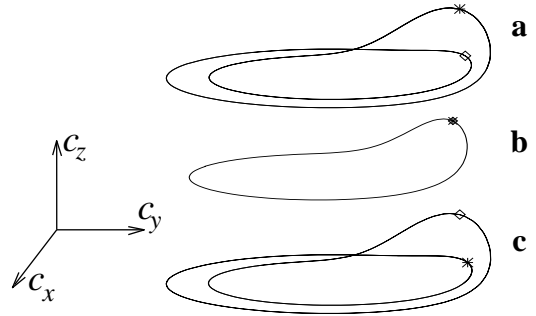}
\end{center}
%\vspace{-1cm}
\caption{Loop exchange in local orbits as the 
$\Omega$ curve is crossed (see text). Data collected on the arc
shown in Fig.~\protect \ref{fig2}:
(a) $\theta_1=\theta_{\Omega}-0.05$; (b) $\theta_2=\theta_{\Omega}$; 
(c) $\theta_3=\theta_{\Omega}+0.05$. Two points corresponding to
time instants $t_1=10.3$ (diamond) and $t_2=19.8$ (asterisk) are marked
on each trajectory to highlight the exchange.}
\label{fig5}
\end{figure}

The nature of the phase jump associated with the $\Omega$ curve 
can be understood from the observation of the local orbit loop exchange 
which occurs on the scale of the $\Omega$ curve width. Figure~\ref{fig5} 
shows local orbits calculated at $r_0=130$ for three angles 
$\theta_1=\theta_{\Omega}-0.05, \theta_2=\theta_{\Omega},
\theta_3=\theta_{\Omega}+0.05$. 
(The corresponding $c_z(t)$ time series for the first and last
orbits are shown in Fig.~\ref{fig4} and $\delta c_z(\theta)$ 
is given in Fig.~\ref{fig3}(c).)
Consider local orbits on the arc connecting points $\theta_1$ and $\theta_3$
(cf. Fig.~\ref{fig2}). As one traverses the arc, the larger, outer loop 
of the local orbit constantly shrinks while the smaller,
inner loop grows. At $\theta=\theta_{\Omega}$, both loops merge  
and then pass each other
exchanging their positions in phase space. 
(Compare Figs.~\ref{fig5}(a) and (c).) The behavior
of the $\delta c_z(\theta)$ order parameter along the arc provides only 
limited information on the loop exchange.  As one sees from Fig.~\ref{fig5}, 
not only are the $c_z$ maximum values of two loops equal 
($\delta c_z(\theta_{\Omega})=0$), but the entire loops 
coincide in phase space: at the exchange point $\theta=\theta_{\Omega}$,
the local oscillation is effectively period-1 \cite{PRE}. 
Continuity of the medium requires that the period-1 points form a line 
extending from the core to the boundary: the $\Omega$ curve.
The $\Omega$ curve separates domains that are dynamically
equivalent apart from a phase shift and it should have no net
velocity in its normal direction. Outside the core region, where
the $\Omega$ curve width is small on the scale of the spiral
wavelength, any large-scale curvature will be eliminated by motion 
of the $\Omega$ curve. This motion, induced by the mean curvature and 
proportional to the diffusion coefficient, 
will yield a straight $\Omega$ curve. In the 
core region, the $\Omega$ curve width is comparable to the length scale of
other concentration gradients and more complicated structure is
possible. \cite{PRE}

\begin{figure}[htbp]
\begin{center}
\leavevmode
\epsfysize 6.5cm
\epsffile{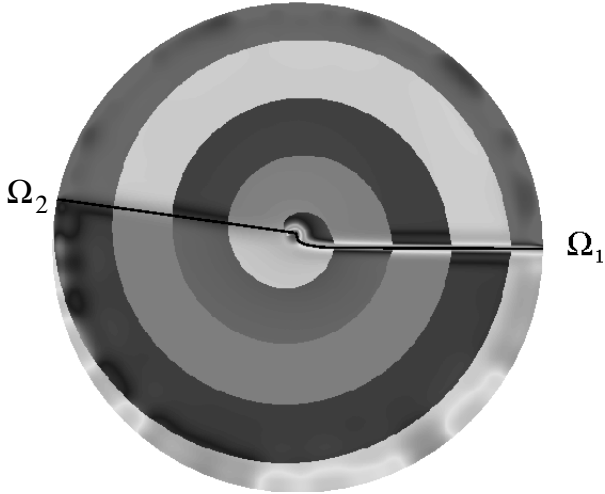}
\end{center}
\caption{Synchronization domains delimited by $\Omega_1$ and $\Omega_2$ curves 
in a medium with period-4 dynamics. The maximum value of $c_z({\bf r})$ in one
spiral rotation is plotted as grey shades. Apart from boundary effects, 
roughly four levels are observed, corresponding to the four loops of the local
orbit, from loop~1 (darkest shade) to loop~4 (lightest shade).}
\label{fig6}
\end{figure}

We now consider more complex local oscillations.
At $C \approx 4.09$, the system undergoes a bifurcation to period 4
which  can be described by an order parameter $\delta_2 c_z({\bf r})$, 
analogous to $\delta c_z({\bf r})$, defined 
as the difference between the first and third or second and fourth 
$c_z({\bf r},t)$ maxima within one full period of the oscillation. 
This can be generalized to any period-$2^k$ regime.
An oscillatory medium with period-4 local dynamics has two types of 
synchronization defect lines: $\Omega_1$ and $\Omega_2$ curves 
with associated phase jump values of $\pm 6\pi$ and $4\pi$, 
respectively \cite{sign}. These curves are shown in Fig.~\ref{fig6}
for $C=4.3$.

We may number the loops of the local period-4 cycles successively 
according to their positions in phase space starting from the innermost loop.
Then the $\Omega_1$ curve corresponds to
the $(^{1234}_{4312})$ loop exchange altering all four loops, while 
the $\Omega_2$ curve is attributed to the $(^{1234}_{2143})$ exchange 
which involves only rearrangement of loops inside the period-2 bands. 
By considering topological braid properties of a period-doubled
orbit \cite{PRE}, all possible loop exchanges can be enumerated.
Outside the core region, we expect that there exist $k$ types of 
$\Omega$ curves in a period-$2^k$ medium. 
The argument given above for the value of the 
$\oint \nabla \phi ({\bf r},t) \cdot d{\bf l}$
integral applies to a period-4 medium as well. 
The existence of two types of defect provides
several ways to satisfy condition (\ref{balance}).
Thus, a medium with a one-armed spiral wave characterized by 
$n_{\rm t} = +1$ may have one $\Omega_1$ curve with $+6\pi$ phase shift 
or both $\Omega_1$ ($-6\pi$) and $\Omega_2$ ($4\pi$) curves.
Configurations where more than two curves originate in the same spiral core
are, in principle, possible. However, they were not observed starting from
the initial conditions considered here. 
 
For higher values of $C$, the $\Omega$ curves may evolve in time leading to 
spatio-temporal chaotic regimes within the general spiral structure.
The curves  separate domains of near synchronization
where the local phase $\phi({\bf r},t)$ changes continuously 
but experiences sudden jumps when a domain boundary is crossed. A full
exploration of the dynamics of $\Omega$ curves in this and other
parameter regimes remains to be carried out. 

Wavelength-doubled spiral waves have been observed in recent
experiments on the Belousov-Zhabotinsky (BZ) reaction in ruthenium-complex 
monolayers on the surface of a BZ solution.~\cite{exp} The
experimentally observed spiral wave structure with a clearly 
visible $\Omega$ curve (cf. Fig.~5 of Ref.~\cite{exp}) is identical 
to that in Fig.~\ref{fig2}. Not only is the final wavelength-doubled 
spiral wave sturcture the same, but the dynamics leading to its
formation from an initial period-1 spiral is also the same as that
observed in the experiment. We have demonstrated that such 
broken-symmetry spiral waves are necessarily formed when the local 
dynamics has period-2 character, as is the case in Ref.~\cite{exp}. 
Although our results were obtained for a system 
possessing a period-doubling cascade, we believe that the 
synchronization defect lines are generic features of distributed media
exhibiting complex periodic behavior, independent of the
specific origin of this periodicity. Thus, we expect the phenomena
described in this Letter to be observable in excitable as well as 
oscillatory media where such period-$n$ local dynamics is observed.

Work supported in part by a grant from the Natural Sciences and
Engineering Research Council of Canada. We would like to thank M.
Golubitsky for discussions on symmetry breaking and period-doubling
bifurcations.

\end{multicols}


\begin{thebibliography}{99}
\bibitem{localized} The localized character of the $\Omega$ curve can be 
understood intuitively as a consequence of the existence of the unstable
orbit separating the two loops of the cycle in the local phase space.
In this respect, the curve is similar to a kink in a bistable system.
\bibitem{sign} The phase shift for the $\Omega_2$ curve is
$4\pi$, i.e., exactly equal to one half of a full-period 
phase increment of $8\pi$. Thus,
it is impossible to distinguish oscillations shifted forward or 
backward by $4\pi$,  
and the phase shift of the $\Omega_2$ curve is left unsigned, 
like that for the $\Omega$ curve for period-2.
\end{thebibliography}
\end{document}